\documentclass[prd,aps,floats,twocolumn]{revtex4}

\usepackage[dvips]{graphicx}
\usepackage{amssymb}
\usepackage{amsmath}
\begin{document}

\title{Probing the early universe with inflationary gravitational
  waves}

\author{Latham A. Boyle and Paul J. Steinhardt}

\affiliation{Department of Physics, Princeton University,
  Princeton, New Jersey 08544, USA \\ }

\date{November 2005}
              
\begin{abstract}
  Near comoving wavenumber $k$, the gravitational-wave background
  (GWB) from inflation carries information about the physical
  conditions near two moments in cosmic history: the moment when $k$
  ``left the horizon'' during inflation, and the moment when it
  ``re-entered the horizon'' after inflation.  We investigate the
  extent to which this information can be extracted if the GWB is
  measured by a combination of cosmic-microwave-background (CMB)
  polarization experiments on large scales and space-based
  laser-interferometer experiments on small scales.  To disentangle
  this information, we derive a new gravitational-wave transfer
  function that incorporates a number of physical effects that were
  treated less accurately, less generally, or were missing altogether
  in previous treatments.  In particular, it incorporates: {\it (i)}
  dark energy with time-varying equation-of-state $w(z)$; {\it (ii)}
  tensor anisotropic stress due to free-streaming relativistic
  particles in the early universe; and {\it (iii)} a variety of
  physical effects that cause deviations from the standard
  equation-of-state $w=1/3$ during the radiation era.  Based on this
  transfer function, we consider the degree to which the GWB can be
  used to test inflation and to probe the ``primordial dark age''
  between the end of inflation and the electroweak phase transition.
\end{abstract}
\maketitle

\section{Introduction}
\label{intro}

Inflation \cite{inflation} generates tensor perturbations
(gravitational waves) with a nearly scale-invariant primordial power
spectrum \cite{early_gwave}.  This gravitational-wave background (GWB)
contributes to the cosmic microwave background (CMB) temperature
anisotropy \cite{early_cmb,cmb_temp}, and also produces a
characteristic curl (or ``B-mode'') fingerprint in the CMB
polarization \cite{cmb_polar,curlmodes,bmodes}.  Several CMB
experiments are being developed to pursue this signal
\cite{cmb_experiments}.  The GWB also persists as a sea of relic
gravitational radiation filling the universe today
\cite{turner_bound,SmithKamCoor}.  Direct detection of this relic
radiation has received considerable attention over the past year or
so, since it has been realized that space-based laser interferometers
operating in the frequency range $0.1~{\rm Hz}<f<1~{\rm Hz}$ might
achieve the necessary sensitivity and foreground subtraction
\cite{DECIGO,Phinney}.  In particular, two futuristic experiments have
been proposed --- NASA's ``Big Bang Observer'' (BBO) and the Japanese
``Deci-hertz Interferometer Gravitational Wave Observatory'' (DECIGO)
--- and are currently being investigated \cite{BBO,Ungarelli,Crowder}.

The gravitational wave spectrum generated by inflation carries
important information about the conditions during inflation.  But the
spectrum also receives corrections, both large and small, from the
subsequent evolution and matter content of the universe after
inflation.  In this paper, we identify various post-inflationary
physical effects which modify the GWB, and show how they may be
encoded in the gravitational-wave transfer function that relates the
primordial tensor power spectrum to the gravitational-wave spectrum at
a later point in cosmic history.  It is necessary to properly
understand and disentangle the post-inflationary effects in order to
optimally extract the inflationary information in the GWB.  But these
modifications are also interesting in their own right, since they
offer a rare window onto the physical properties of the high-energy
universe during the ``primordial dark age'' between the end of
inflation and the electroweak phase transition.

The same features that make the inflationary GWB difficult to detect
--- namely its small amplitude and gravitational-strength coupling to
matter --- also make it a clean cosmological probe.  First, because of
their tiny amplitude, the gravitational waves obey linear equations of
motion, so that their evolution may be predicted analytically with
high precision.  (By contrast, density perturbations grow after
horizon entry, and perturbations shorter than $\sim10$~Mpc have
already grown non-linear; so it is impossible to study their evolution
analytically, and even numerical analysis is challenging.)  Second,
because of their ultra-weak interactions with matter, the
gravitational waves have been free streaming since the end of
inflation --- in contrast to neutrinos (which began streaming roughly
a second later) and photons (which began streaming several hundred
thousand years later). The gravitational waves carry unsullied
information from the early universe, and subsequent modifications of
the gravitational-wave spectrum are not washed out by thermal effects
(since the gravitons are thermally decoupled).

The gravitational-wave spectrum near a given wavenumber $k$ is
primarily sensitive to two ``moments'' in cosmic history: (1) the
moment when that mode ``left the horizon'' ({\it i.e.}, became longer
than the instantaneous Hubble radius during inflation), and (2) the
moment when the mode ``re-entered the horizon'' ({\it i.e.}, became
shorter than the instantaneous Hubble radius once again, after the end
of inflation).  The first moment imprints information about inflation
itself, while the second moment imprints information about
post-inflationary conditions.  The CMB is sensitive to long-wavelength
modes that re-entered at relatively low temperatures (well after BBN),
corresponding to relatively well-understood physics.  By contrast,
laser interferometers are sensitive to shorter wavelengths that
entered the horizon at high temperatures ($T\sim10^{7}$~GeV), well
above the electroweak phase transition.  The physical conditions at
such high energies, which are considerably beyond the reach of
particle accelerators, are a mystery, so that any information about
this epoch from the GWB would be very valuable.

The outline of our paper is as follows.  In section
\ref{fundamentals}, we review the basics of inflationary gravitational
waves.  We take special care to clearly establish our conventions, and
to point out where conventions diverge and become confused in the
literature.  In section \ref{transfer_fn} we present a new calculation
of the gravitational-wave transfer function which includes effects not
considered in previous calculations
\cite{SmithKamCoor,TurnLidsWhite,Bashinsky}.  We separate the transfer
function into three factors, each with a distinct physical meaning.

The first factor accounts for the redshift-suppression of the
gravitational-wave amplitude after horizon re-entry.  Among other
things, this factor accomodates a dark energy component with a
time-varying equation-of-state parameter $w(z)$.  The second factor
captures the behavior of the background equation-of-state parameter
$w$ near the time of horizon re-entry.  The third factor accounts for
the damping of tensor modes due to tensor anisotropic stress from
free-streaming relativistic particles in the early universe.  This
damping effect was recently pointed out by Weinberg \cite{Weinberg},
and the damping on CMB scales due to free-streaming neutrinos has been
studied in \cite{Weinberg,PritchKam,Bashinsky,DicusRepko}.  In this
paper, we stress that it is also necessary to consider this damping
effect on laser-interferometer scales, which re-entered the horizon
when free-streaming particles were an unknown fraction $f$ of the
background energy density.  We present accurate expressions for the
damping effect as a function of $f$.  We also observe that Weinberg's
analysis, which originally focused on a single fermionic particle (the
neutrino), extends in a simple way to the more general case of a
mixture of free-streaming bosons and fermions with different
temperatures and decoupling times.

We identify six physical effects which can modify the relic GWB by
causing the equation-of-state parameter $w$ to deviate from its
standard value ($w=1/3$) during the radiation-dominated epoch.
Furthermore, although it is often treated as a stationary random
process, the inflationary GWB is actually highly non-stationary (as
emphasized by Grishchuk \cite{GrishchukSidorov}; also see
\cite{AllenFlanaganPapa,Albrecht}).  Thus, our transfer function keeps
track of the coherent phase information that it contains.  Finally, in
section \ref{discussion}, we discuss the implications of these results
for future CMB and laser-interferometer experiments, and consider what
we might learn by measuring the GWB in both ways.

\section{Tensor perturbations: fundamentals and conventions}
\label{fundamentals}

In this section, we derive some basic facts about inflationary
gravitational waves.  These derivations are not new, but are intended
to make the present paper more pedagogical and self-contained.  They
also establish our conventions explicitly, and provide a brief guide
to conventions used by other authors.  Such a guide is necessary since
there are many slightly different conventions floating around in the
inflationary gravitational-wave literature and, as a result, erroneous
factors of $2$ and $\pi$ are ubiquitous.  We have been careful to
highlight each spot where convention choices crop up, and to state
which convention we have chosen.

Tensor perturbations in a spatially-flat Friedmann-Robertson-Walker
(FRW) universe are described by the line element
\begin{equation}
  {\rm d}s^{2}=a^{2}[-{\rm d}\tau^{2}+(\delta_{ij}+h_{ij})
  {\rm d}x^{i}{\rm d}x^{j}],
\end{equation}
where $\tau$ is the conformal time, $x^{i}$ are comoving spatial
coordinates, and $h_{ij}$ is the gauge-invariant tensor metric
perturbation.  Note that, although our definition of $h_{ij}$ is
perhaps the most common (see {\it e.g.}\ 
\cite{Weinberg,MFB,Maldacena}), some authors define the tensor
perturbation with an extra factor of $2$ (see {\it e.g.}\ 
\cite{Bardeen,KodamaSasaki}) so that the spatial metric is written
$\delta_{ij}+2h_{ij}$.  Throughout this appendix, if a term contains a
repeated index, that index should be summed: from 1 to 3 for latin
indices and from 0 to 3 for greek indices.  The perturbation $h_{ij}$
is symmetric ($h_{ij}\!=\!h_{ji}$), traceless ($h_{ii}\!=\!0$), and
transverse ($h_{ij,j}\!=\!0$), and therefore contains
$6\!-\!1\!-\!3\!=\!2$ independent modes (corresponding to the ``$+$''
and ``$\times$'' gravitational-wave polarizations).

One can think of $h_{ij}(\tau,{\bf x})$ as a quantum field in an
unperturbed FRW background metric $\bar{g}_{\mu\nu}
=diag\{-a^{2},a^{2},a^{2},a^{2}\}$.  At quadratic order in $h_{ij}$
(which is adequate, since $h_{ij}$ is tiny), tensor perturbations are
governed by the action (see {\it e.g.}\ \cite{MFB,Maldacena})
\begin{equation}
  \label{tensor_action}
  S=\int d\tau d{\bf x}\sqrt{-\bar{g}}\left[
    \frac{-\bar{g}^{\mu\nu}}{64\pi G}\partial_{\mu}h_{ij}
    \partial_{\nu}h_{ij}+\frac{1}{2}\Pi_{ij}h_{ij}\right].
\end{equation}
where $\bar{g}^{\mu\nu}$ and $\bar{g}$ are the inverse and determinant
of $\bar{g}_{\mu\nu}$, respectively, and $G$ is Newton's constant.
The tensor part of the anisotropic stress
\begin{equation}
  \Pi_{ij}=T^{i}_{j}-p\delta^{i}_{j}
  \quad(p={\rm unperturbed}\;{\rm pressure})
\end{equation}
satisfies $\Pi_{ii}=0$ and $\partial_{i}\Pi_{ij}=0$, and couples to
$h_{ij}$ like an external source in (\ref{tensor_action}).  By varying
$h_{ij}$ in (\ref{tensor_action}), we obtain the equation of motion
\begin{equation}
  \label{h_eq}
  h_{ij}''+2\frac{a'(\tau)}{a(\tau)}h_{ij}'-{\bf \nabla}^{2}h_{ij}
  =16\pi G a^{2}(\tau)\Pi_{ij}(\tau,{\bf x}),
\end{equation}
where a prime ($\,'\,$) indicates a conformal time derivative
$d/d\tau$.  Next, it is convenient to Fourier transform as follows:
\begin{subequations}
  \label{fourier_expand}
  \begin{eqnarray}
    h_{ij}^{}(\tau,{\bf x})\!\!&=\!&\!\!\sum_{r}
    \sqrt{16\pi G}\!\!\int\!\!\!
    \frac{d{\bf k}}{(2\pi)^{3/2}}\epsilon_{ij}^{r}({\bf k})
    h_{{\bf k}}^{r}(\tau){\rm e}^{i{\bf k}{\bf x}},\qquad\quad \\
    \Pi_{ij}^{}(\tau,{\bf x})\!\!&=\!&\!\!\sum_{r}
    \sqrt{16\pi G}\!\!\int\!\!\!
    \frac{d{\bf k}}{(2\pi)^{3/2}}\epsilon_{ij}^{r}({\bf k})
    \Pi_{{\bf k}}^{r}(\tau){\rm e}^{i{\bf k}{\bf x}},\qquad\quad 
  \end{eqnarray}
\end{subequations}
where $r=$(``$+$'' or ``$\times$'') labels the polarization state, and
the polarization tensors are symmetric [$\epsilon_{ij}^{r} ({\bf
  k})=\epsilon_{ji}^{r}({\bf k})$], traceless [$\epsilon_{ii}^{r}
({\bf k})=0$], and transverse [$k_{i}\epsilon_{ij}^{r}({\bf k})=0$].
We also choose a circular-polarization basis in which
$\epsilon_{ij}^{r}({\bf k})=(\epsilon_{ij}^{r} (\textrm{-}{\bf
  k}))^{\ast}$, and normalize the polarization basis as follows:
\begin{equation}
  \label{basis_norm}
  \sum_{i,j}\epsilon_{ij}^{r}({\bf k})(\epsilon_{ij}^{s}
  ({\bf k}))^{\ast}=2\delta^{rs}.
\end{equation}
Although our normalization convention (\ref{basis_norm}) is the most
standard one, other conventions --- {\it i.e.}\ different numerical
constants on the right-hand side of (\ref{basis_norm}) --- exist in
the literature.  Substituting (\ref{fourier_expand}) into
(\ref{tensor_action}) then yields
\begin{equation}
  \label{tensor_action_fourier}
  S\!=\!\!\sum_{r}\!\!\int\!\!d\tau d{\bf k}\frac{a^{2}}{2}\;\!\!
  \Big[h_{{\bf k}}^{r}{}'h_{\!\textrm{-}{\bf k}}^{r}\!{}'\!
  -\!k^{2}h_{{\bf k}}^{r}h_{\!\textrm{-}{\bf k}}^{r}\! 
  +\!32\pi G a^{2}\Pi_{{\bf k}}^{r}h_{\!\textrm{-}{\bf k}}^{r}
  \Big].
\end{equation}

Now we can canonically quantize by promoting $h_{{\bf k}}^{r}$ and its
conjugate momentum
\begin{equation}
  \label{def_pi}
  \pi_{{\bf k}}^{r}(\tau)=a^{2}(\tau)h_{\!\textrm{-}{\bf k}}^{r}{}'(\tau)
\end{equation}
to operators, $\hat{h}_{{\bf k}}^{r}$ and $\hat{\pi}_{{\bf k}}^{r}$,
satisfying the equal-time commutation relations
\begin{subequations}
  \label{h_pi_commutators}
  \begin{eqnarray}
    \left[\hat{h}_{{\bf k}}^{r}(\tau),
      \hat{\pi}_{{\bf k}'}^{s}(\tau)\right]
    &=&i\delta^{rs}\delta^{(3)}({\bf k}-{\bf k}'), \\
    \left[\hat{h}_{{\bf k}}^{r}(\tau),
      \hat{h}_{{\bf k}'}^{s}(\tau)\right]
    &=&\left[\hat{\pi}_{{\bf k}}^{r}(\tau),
      \hat{\pi}_{{\bf k}'}^{s}(\tau)\right]=0.
  \end{eqnarray}
\end{subequations}
Since $\hat{h}_{ij}(\tau,{\bf x})$ is hermitian, its fourier
components satisfy $\hat{h}_{{\bf k}}^{r}=\hat{h}_{\!\textrm{-}{\bf
    k}}^{r\dag}$, and we write them as
\begin{equation}
  \label{h_from_a}
  \hat{h}_{{\bf k}}^{r}(\tau)=h_{k}^{}(\tau)\hat{a}_{{\bf k}}^{r}
  +h_{k}^{\ast}(\tau)\hat{a}_{\!\textrm{-}{\bf k}}^{r\dag},
\end{equation}
where the creation and annihilation operators,
$\hat{a}_{{\bf k}}^{r\dag}$ and $\hat{a}_{{\bf k}}^{r}$, satisfy
standard commutation relations
\begin{subequations}
  \label{a_commutators}
  \begin{eqnarray}
    \Big[\hat{a}_{{\bf k}}^{r},\hat{a}_{{\bf k}'}^{s\dag}\Big]
    &=&\delta^{rs}\delta^{(3)}({\bf k}-{\bf k}'), \\
    \Big[\hat{a}_{{\bf k}}^{r},\hat{a}_{{\bf k}'}^{s}\Big]
    &=&\Big[\hat{a}_{{\bf k}}^{r\dag},\hat{a}_{{\bf k}'}^{s\dag}\Big]=0,
  \end{eqnarray}
\end{subequations}
while the ($c$-number) mode functions $h_{k}(\tau)$ and
$h_{k}^{\ast}(\tau)$ are linearly-independent solutions of the
fourier-transformed equation of motion
\begin{equation}
  \label{h_eq_ft}
  h_{k}''+2\frac{a'(\tau)}{a(\tau)}h_{k}'+k^{2}h_{k}^{}=
    16\pi G a^{2}(\tau)\Pi_{k}^{}(\tau).
\end{equation}
Eq.\ (\ref{h_from_a}) makes use of the fact that, by isotropy, the
mode functions $h_{k}^{}(\tau)$ will depend on the time ($\tau$) and
the wavenumber ($k=|{\bf k}|$), but not on the direction ($\hat{{\bf
    k}}$) or the polarization ($r$).  Note that consistency between
the two sets of commutation relations, (\ref{h_pi_commutators}) and
(\ref{a_commutators}), requires that the mode functions satisfy the
Wronskian normalization condition
\begin{equation}
  \label{Wronskian}
  h_{k}^{}(\tau)h_{k}^{\ast}{}'(\tau)-h_{k}^{\ast}(\tau)h_{k}'(\tau)
  =\frac{i}{a^{2}(\tau)}
\end{equation}
in the past.  In particular, the standard initial condition for the
mode function in the far past (when the mode $k$ was still far inside
the horizon during inflation):
\begin{equation}
  \label{h_bc}
  h_{k}^{}(\tau)\to\frac{{\rm exp}(-ik\tau)}{a(\tau)\sqrt{2k}}
  \qquad({\rm as}\;\;\tau\to-\infty),
\end{equation}
satisfies (\ref{Wronskian}) --- but it is not the unique initial
condition which does so.  This is a manifestation of the well known
vacuum ambiguity that is responsible for particle production in
cosmological spacetimes (see \cite{BirrellDavies}).

Now that we have discussed the quantization of tensor perturbations,
let us turn to the three different spectra that are commonly used to
describe the stochastic GWB: the tensor power spectrum
$\Delta_{h}^{2}(k,\tau)$, the chirp amplitude $h_{c}^{}(k,\tau)$, and
the energy spectrum $\Omega_{gw}^{}(k,\tau)$.

In the early universe, the GWB is usually characterized by the tensor
power spectrum $\Delta_{h}^{2}(k,\tau)$.  With the formalism developed
thus far, one can check that
\begin{equation}
  \langle0|\hat{h}_{ij}^{}(\tau,{\bf x})\hat{h}_{ij}^{}
  (\tau,{\bf x})|0\rangle\!=\!\!\!\int_{0}^{\infty}\!\!\!\!\!64\pi G
  \frac{k^{3}}{2\pi^{2}}\!\left|h_{k}^{}(\tau)\right|^{2}
  \!\frac{dk}{k},
\end{equation}
so that the tensor power spectrum is given by
\begin{equation}
  \label{def_tensor_power}
  \Delta_{h}^{2}(k,\tau)\equiv\frac{d\langle0|\hat{h}_{ij}^{2}
    |0\rangle}{d\,{\rm ln}\,k}=64\pi G\frac{k^{3}}{2\pi^{2}}
  \left|h_{k}^{}(\tau)\right|^{2}.
\end{equation}
Note that our definition (\ref{def_tensor_power}) agrees with the WMAP
convention (see \cite{Peiris}) --- this will be clearer when we
present the approximate slow-roll form of the spectrum below.
Although the WMAP convention seems to be becoming the standard one,
several other definitions of the tensor power spectrum exist in the
literature, and differ from (\ref{def_tensor_power}) by an overall
numerical constant.  Also, since (\ref{def_tensor_power}) defines the
tensor power spectrum in terms of the full tensor perturbation
$h_{ij}$, the normalization of the power spectrum is independent of
the normalization (\ref{basis_norm}) of the polarization basis.  By
contrast, some authors define the tensor power spectrum in terms of
the polarization components of $h_{ij}$, so that the normalization of
the spectrum is linked to the convention-dependent coefficient on the
right-hand side of (\ref{basis_norm}).

The present-day GWB is usually characterized either by its chirp
amplitude $h_{c}^{}(k,\tau)$, or by its energy spectrum
$\Omega_{gw}^{}(k,\tau)$.  First, the chirp amplitude represents the
rms dimensionless strain ($\sim\delta L/L$ in a gravitational wave
antenna) per logarithmic wavenumber interval (or logarithmic frequency
interval), and is related to the tensor power spectrum as follows:
$h_{c}^{}(k,\tau) =\sqrt{\Delta_{t}^{2}(k,\tau)/2}$ (see \cite{Thorne}
for more details).  The energy spectrum
\begin{equation}
  \label{def_Omega_gw}
  \Omega_{gw}^{}(k,\tau)\equiv\frac{1}{\rho_{crit}^{}(\tau)}
  \frac{d\langle 0|\hat{\rho}_{gw}^{}(\tau)|0\rangle}{d\,{\rm ln}\,k}
\end{equation}
represents the gravitational-wave energy density ($\rho_{gw}^{}$) per
logarithmic wavenumber interval, in units of the ``critical density''
\begin{equation}
  \label{def_rho_crit}
  \rho_{crit}^{}(\tau)=\frac{3H^{2}(\tau)}{8\pi G}.
\end{equation}
To compute $\Omega_{gw}^{}(k,\tau)$, we can again think of the tensor
perturbation $h_{ij}$ as a field in an unperturbed FRW
background metric $\bar{g}_{\mu\nu}$, and then use its action
(\ref{tensor_action}) to compute its stress-energy tensor
\begin{equation}
  \label{T_alpha_beta}
  T_{\alpha\beta}=-2\frac{\delta L}{\delta\bar{g}^{\alpha\beta}}
  +\bar{g}_{\alpha\beta}L,
\end{equation}
where $L$ is the Lagrangian function in (\ref{tensor_action}) --- see,
for example, Ch.~21.3 in \cite{MTW}.  In particular, the
gravitational-wave energy density (ignoring the anisotropic stress
coupling) is
\begin{equation}
  \label{rho_gw}
  \rho_{gw}^{}=-T_{0}^{0}=\frac{1}{64\pi G}
  \frac{(h_{ij}')^{2}+(\vec{{\bf \nabla}}h_{ij})^{2}}{a^{2}},
\end{equation}
which has vacuum expectation value
\begin{equation}
  \label{rho_gw_expect}
  \langle0|\rho_{gw}^{}|0\rangle=\int_{0}^{\infty}\frac{k^{3}}
  {2\pi^{2}}\frac{\left|h_{k}'\right|^{2}
    +k^{2}\left|h_{k}^{}\right|^{2}}{a^{2}}\frac{dk}{k},
\end{equation}
so that the energy spectrum is given by
\begin{equation}
  \Omega_{gw}^{}(k,\tau)=\frac{8\pi G}{3H^{2}(\tau)}
  \frac{k^{3}}{2\pi^{2}}
  \frac{\left|h_{k}'(\tau)\right|^{2}
    +k^{2}\left|h_{k}^{}(\tau)\right|^{2}}{a^{2}(\tau)}.
\end{equation}
We have shown how each of the three spectra ($\Delta_{t}^{2}$,
$h_{c}^{}$, and $\Omega_{gw}^{}$) may be written in terms of the mode
functions $h_{k}^{}(\tau)$.  Also note that, once the mode $k$
re-enters the horizon after inflation, the corresponding mode function
obeys $|h_{k}'(\tau)|^{2}=k^{2}|h_{k}^{}(\tau)|^{2}$, as explained in
the next section, so that the three spectra may be related to one
another in a simple way:
\begin{equation}
  \label{spec_relations}
  \Omega_{gw}(k,\tau)=\frac{1}{12}\frac{k^{2}\Delta_{h}^{2}(k,\tau)}
  {a^{2}(\tau)H^{2}(\tau)}=\frac{1}{6}\frac{k^{2}h_{c}^{2}(k,\tau)}
  {a^{2}(\tau)H^{2}(\tau)}.
\end{equation}

Next, let us ``derive'' the slow-roll expression for the primordial
tensor power spectrum.  As long as $k$ remains inside the Hubble
horizon ($k\gg aH$) during inflation, the mode function $h_{k}^{({\rm
    in})}(\tau)$ is given by (\ref{h_bc}); and once $k$ is outside the
horizon ($k\ll aH$), the mode function $h_{k}^{({\rm out})}$ is
independent of $\tau$.  Then, by simply matching $|h_{k}^{({\rm
    in})}|$ to $|h_{k}^{({\rm out})}|$ at the moment of horizon-exit
($k=aH$), one finds $h_{k}^{({\rm out})} =1/(a_{\ast}^{}\sqrt{2k})$,
where an asterisk ($\ast$) denotes that a quantity is evaluated at
horizon-exit ($k=a_{\ast}^{}H_{\ast}^{}$).  Thus, the primordial
tensor power spectrum is given by
\begin{equation}
  \label{tensor_power_slowroll}
  \Delta_{h}^{2}(k)\equiv64\pi G\frac{k^{3}}{2\pi^{2}}\left|
    h_{k}^{({\rm out})}\right|^{2}\approx8\left(\frac{H_{\ast}}
    {2\pi M_{pl}}\right)^{2},
\end{equation}
where, in this equation (and in the remainder of this section)
``$\equiv$'' denotes a definition, ``$\approx$'' indicates that the
slow-roll approximation has been used, and $M_{pl}=(8\pi G)^{-1/2}$ is
the ``reduced Planck mass.''  Note that the primordial power spectrum
is time-independent, since (by definition) it is evaluated when the
mode $k$ is far outside the horizon.  Although our derivation of
(\ref{tensor_power_slowroll}) may seem crude, it is well known that
(\ref{tensor_power_slowroll}) provides a very good approximation to
the inflationary tensor spectrum.  We will not reproduce it here, but
a closely analogous derivation leads to the slow-roll expression for
the primordial {\it scalar} power spectrum:
\begin{equation}
  \label{scalar_power_slowroll}
  \Delta_{{\cal R}}^{2}(k)\approx\frac{1}{2\epsilon_{\ast}^{}}\left(
    \frac{H_{\ast}^{}}{2\pi M_{pl}^{}}\right)^{2}.
\end{equation}
It is also useful to define the tensor/scalar ratio
\begin{equation}
  \label{r}
  r(k)\equiv\Delta_{h}^{2}(k)/\Delta_{{\cal R}}^{2}(k)
  =16\epsilon_{\ast}^{},
\end{equation}
along with the scalar and tensor spectral indices
\begin{subequations}
  \label{indices}
  \begin{eqnarray}
    \label{ns}
    n_{s}^{}(k)-1&\equiv&\frac{d\,{\rm ln}\,\Delta_{{\cal R}}^{2}}
    {d\,{\rm ln}\,k}\approx-6\epsilon_{\ast}^{}+2\eta_{\ast}^{}, \\
    \label{nt}
    n_{t}^{}(k)&\equiv&\frac{d\,{\rm ln}\,\Delta_{\;\!h\;\!}^{2}}
    {d\,{\rm ln}\,k}\approx-2\epsilon_{\ast}^{}.
  \end{eqnarray}
\end{subequations}
In the slow-roll formulae above, we have introduced the usual ``potential''
slow roll parameters
\begin{equation}
  \label{def_eps_eta}
  \epsilon\equiv\frac{1}{2}M_{pl}^{2}\left(\frac{V'(\phi)}{V(\phi)}
    \right)^{2},\qquad\eta\equiv M_{pl}^{2}\frac{V''(\phi)}{V(\phi)},
\end{equation}
where $V(\phi)$ is the inflaton potential.  Note that Eqs.\ (\ref{r})
and (\ref{nt}) together imply the well known consistency relation for
inflation
\begin{equation}
  \label{consistency}
  r=-8n_{t}^{}.
\end{equation}

\section{The tensor transfer function}
\label{transfer_fn}

Since cosmological tensor perturbations are tiny, they are well
described by linear perturbation theory, so that each fourier mode
evolves independently.  Thus, we see from Eq.\ 
(\ref{def_tensor_power}) that the primordial tensor power spectrum ---
defined at some conformal time $\tau_{i}^{}$ shortly after the end of
inflation, when all modes of interest have already left the horizon,
but have not yet re-entered --- is related to the tensor power
spectrum at a later time $\tau$ by a multiplicative transfer function
\begin{equation}
  \label{def_Tk}
  \Delta_{h}^{2}(k,\tau)=T_{h}^{}(k,\tau)\Delta_{h}^{2}(k,\tau_{i}^{}),
\end{equation}
where
\begin{equation}
  \label{Th}
  T_{h}^{}(k,\tau)=\left|\frac{h_{k}(\tau)}{h_{k}(\tau_{i}^{})}
  \right|^{2}.
\end{equation}
Note that we will not necessarily want to evaluate $T_{h}^{}(k,\tau)$
at the present time ($\tau=\tau_{0}^{}$), since different experiments
probe the gravitational-wave spectrum at different redshifts.  For
example, while laser interferometers measure $T_{h}^{}$ today, CMB
experiments measure it near the redshift of recombination.  As long as
a mode remains outside the horizon ($k\!\ll\!aH$), the corresponding
perturbation does not vary with time [$h_{k}^{}(\tau)\!=\!h_{k}^{}
(\tau_{i})$], so that the transfer function is very well approximated
by $T_{h}^{}(k,\tau)=1$.  (For a general proof, even in the presence
of anisotropic stress, see the appendix in \cite{Weinberg}.)  Thus,
the rest of this paper will focus on $T_{h}^{}(k,\tau)$ for modes that
have already re-entered the horizon prior to time $\tau$.

It is very convenient to split the transfer function (\ref{Th}) into
three factors as follows:
\begin{equation}
  T_{h}^{}(k,\tau)=\left|\frac{\bar{h}_{k}(\tau)}{h_{k}(\tau_{i}^{})}
    \frac{\widetilde{h}_{k}(\tau)}{\bar{h}_{k}(\tau)}
    \frac{h_{k}^{}(\tau)}{\widetilde{h}_{k}(\tau)}\right|^{2}
  =C_{1}C_{2}C_{3}.
\end{equation}
Here $h_{k}(\tau)$, $\widetilde{h}_{k}(\tau)$ and $\bar{h}_{k}(\tau)$
represent three different solutions of the tensor mode equation
(\ref{h_eq_ft}).  In particular, $h_{k}(\tau)$ is the true (exact)
solution of (\ref{h_eq_ft}); $\widetilde{h}_{k}(\tau)$ is an
approximate solution obtained by ignoring the tensor anisotropic
stress $\Pi_{k}$ on the right-hand-side of (\ref{h_eq_ft}); and
$\bar{h}_{k}(\tau)$ is an even cruder approximation obtained by first
ignoring $\Pi_{k}$ and then using the ``thin-horizon'' approximation,
described in subsection \ref{thin_horizon}, to solve (\ref{h_eq_ft}).
[Briefly, the thin-horizon approximation treats horizon re-entry as a
``sudden'' or instantaneous event.  In this approximation,
$\bar{h}_{k}(\tau)$ is frozen outside the Hubble horizon, redshifts as
$1/a(\tau)$ inside the Hubble horizon, and has a sharp transition
between these two behaviors at the moment when the mode re-enters the
Hubble horizon ($k=aH$).]

These three factors each represent a different physical effect.  The
first factor,
\begin{equation}
  \label{def_C1}
  C_{1}=\left|\frac{\bar{h}_{k}(\tau)}{h_{k}(\tau_{\!i}^{})\!}\right|^{2},
\end{equation}
represents the redshift-suppression of the gravitational-wave
amplitude after the mode $k$ re-enters the horizon.  The second
factor,
\begin{equation}
  \label{def_C2}
  C_{2}=\left|\frac{\widetilde{h}_{k}(\tau)}{\bar{h}_{k}(\tau)}\right|^{2},
\end{equation}
captures the behavior of the background equation-of-state parameter
$w(\tau)=p(\tau)/\rho(\tau)$ around the time of horizon re-entry.  The
third factor,
\begin{equation}
  \label{def_C3}
  C_{3}=\left|\frac{h_{k}(\tau)}{\widetilde{h}_{k}(\tau)}\right|^{2},
\end{equation}
measures the damping of the gravitational-wave spectrum due to tensor
anisotropic stress.  Note that $C_{1}$ by itself is $\ll 1$ and
provides a rough approximation to the full transfer function $T_{h}$.
The other two factors, $C_{2}$ and $C_{3}$, are typically of order
unity, and are primarily sensitive to the physical conditions near the
time that the mode $k$ re-entered the Hubble horizon.  In subsections
\ref{thin_horizon}, \ref{horizon_thickness}, and \ref{free_streaming},
we derive expressions for $C_{1}$, $C_{2}$ and $C_{3}$, respectively,
and flesh out the physical interpretations given above.  In subsection
\ref{EOS_corrections}, we discuss various physical effects that cause
deviations $\delta w$ from the usual equation-of-state parameter
($w=1/3$) in the early universe, and explain how these effects modify
the gravitational-wave transfer function.

\subsection{The redshift-suppression factor, $C_{1}$}
\label{thin_horizon}

The mode function $h_{k}^{}(\tau)$ behaves simply in two regimes: far
outside the horizon ($k\ll aH$), and far inside the horizon ($k\gg
aH$).  Far outside, $h_{k}^{}(\tau)$ is $\tau$-independent (as we have
seen).  Far inside, after horizon re-entry, $h_{k}^{}(\tau)$
oscillates with a decaying envelope
\begin{equation}
  \label{hk_inside}
  h_{k}^{}(\tau)=\frac{A_{k}^{}}{a(\tau)}
  {\rm cos}[k(\tau-\tau_{k}^{})+\phi_{k}^{}],
\end{equation}
as we shall see in the next two subsections, where $A_{k}^{}$ and
$\phi_{k}^{}$ are constants representing the amplitude and phase shift
of the oscillation, and $\tau_{k}^{}$ is the conformal time at horizon
re-entry ($k=aH$).  These two simple regimes are separated by an
intermediate period (horizon-crossing) when $k\sim aH$.

In the thin-horizon approximation, we neglect this intermediate
regime.  That is, we assume that $\bar{h}_{k}^{}(\tau)=h_{k}^{}
(\tau_{i}^{})$ when $k<aH$; and that $\bar{h}_{k}^{}(\tau)$ is given
by Eq.\ (\ref{hk_inside}) for $k>aH$; and that the outside amplitude
is connected to the inside envelope via the matching condition
$h_{k}^{}(\tau_{i})=A_{k}/a(\tau_{k})$.  Ignoring the phase shift
$\phi_{k}$, which is really an asymptotic quantity, this matching
condition simply imposes continuity of the inside and outside
amplitudes at $k=aH$.  Combining the matching condition with Eq.\ 
(\ref{def_C1}), we see that
\begin{equation}
  \label{C1}
  C_{1}=\left(\frac{1+z\;}{1+z_{k}^{}}\right)^{2}
  {\rm cos}^{2}[k(\tau-\tau_{k}^{})+\phi_{k}^{}],
\end{equation}
where $1+z=a_{0}^{}/a(\tau)$ is the redshift at which the spectrum is
to be probed, and $1+z_{k}^{}=a_{0}^{}/a_{k}^{}$ is the redshift at
which the mode re-entered the Hubble horizon ($k=aH$).

The relic GWB from inflation is often treated as a
``quasi-stationary'' process (which means that its statistical
properties only vary on cosmological time scales --- much longer than
the timescales in a terrestrial experiment).  But the ${\rm cos}^{2}
[\ldots]$ factor in Eq.\ (\ref{C1}) implies that the background is
actually highly {\it non-stationary} --- its power spectrum oscillates
as a function of both wavenumber $k$ and time $\tau$.  This ${\rm
  cos}^{2}[\ldots]$ factor represents a genuine feature, and is {\it
  not} a spurious byproduct of our thin-horizon approximation.
Physically (as observed in \cite{GrishchukSidorov}) the inflationary
GWB consists of gravitational {\it standing waves} with random {\it
  spatial} phases, and coherent {\it temporal} phases.  All modes
$\vec{k}$ at fixed wavenumber $k=|\vec{k}|$ re-enter the Hubble
horizon simultaneously, and subsequently oscillate in phase with one
another --- even until the present day.  Classically, the modes are
synchronized by inflation; quantum mechanically, they are squeezed
\cite{GrishchukSidorov}.  Thus, at a fixed wavenumber, $T_{h}(k,\tau)$
is sinusoidal in $\tau$, with oscillation frequency $k$, and a phase
shift $\phi_{k}^{}$ (computed in the next subsection).  Alternatively,
at fixed time, $T_{h}(k,\tau)$ oscillates rapidly in wavenumber.  For
the direct detection experiments we are considering, which cannot
resolve these oscillations, the factor ${\rm cos}^{2}[\ldots]$ should
be averaged, and replaced by $1/2$ in Eq.\ (\ref{C1}).

In the remainder of this subsection, we derive an accurate expression
for $(1+z_{k}^{})$, the horizon-crossing redshift.  To start, let us
write the background energy density $\rho$ as a sum of several
components.  The $i$th component has energy density $\rho_{i}^{}$,
pressure $p_{i}^{}$, equation-of-state parameter $w_{i}^{}\equiv
p_{i}^{}/\rho_{i}^{}$, and obeys a continuity equation
\begin{equation}
  \label{continuity}
  \frac{d\rho_{i}^{}}{\rho_{i}^{}}=3[1+w_{i}^{}(z)]\frac{dz}{1+z}.
\end{equation}
Integrating this equation yields
\begin{equation}
  \rho_{i}^{}(z)/\rho_{i}^{(0)}=(1+z)^{3}{\rm exp}\!
  \left[3\!\int_{0}^{z}\frac{w_{i}^{}(\tilde{z})}
    {1+\tilde{z}}d\tilde{z}\right],
\end{equation}
where $\rho_{i}^{(0)}$ is the present value.  Then the Friedmann
equation
\begin{equation}
  H^{2}(z)=\frac{8\pi G}{3}\sum_{i}\rho_{i}^{}(z)
\end{equation}
may be rewritten as
\begin{equation}
  \frac{a^{2}H^{2}}{a_{0}^{2}H_{0}^{2}}
  =\sum_{i}\Omega_{i}^{(0)}(1+z){\rm exp}\!\left[3\!\int_{0}^{z}
    \frac{w_{i}^{}(\tilde{z})}{1+\tilde{z}}d\tilde{z}\right],
\end{equation}
where $H_{0}^{}$ is the present Hubble rate, and the density parameter
$\Omega_{i}^{(0)}\equiv\rho_{i}^{(0)}\!/\rho_{cr}^{(0)}$ represents
the $i$th component's present energy density in units of the present
critical density $\rho_{cr}^{(0)}= 3H_{0}^{2}/8\pi G$.  Hence
$z_{k}^{}$ is obtained by solving the equation
\begin{equation}
  \label{zk_Friedmann}
  (k/k_{0}^{})^{2}=F(z_{k}^{})
\end{equation}
where 
\begin{equation}
  \label{def_F}
  F(z_{k}^{})=\sum_{i}\Omega_{i}^{(0)}(1+z_{k}^{}){\rm exp}\!
  \left[3\!\int_{0}^{z_{k}^{}}\frac{w_{i}^{}(z)}{1+z}dz\right],
\end{equation}
and $k_{0}^{}=a_{0}^{}H_{0}^{}=h\times 3.24\times 10^{-18}\,{\rm Hz}$
is today's horizon wavenumber.  

Before solving this equation properly, let us pause to extract a few
familiar approximate scalings from our formalism.  Since the
primordial inflationary power spectrum $\Delta_{h}^{2}(k,\tau_{i})$ is
roughly scale invariant [$\propto(k/k_{0}^{})^{0}$], the current power
spectrum $\Delta_{h}^{2}(k,\tau_{0}^{})$ is roughly $\propto C_{1}$,
and hence $\propto(1+z_{k}^{})^{-2}$.  From Eq.\ 
(\ref{spec_relations}) we have $h_{c}^{}(k,\tau_{0}^{})\propto
(1+z_{k}^{})^{-1}$ and $\Omega_{gw}^{}(k,\tau_{0}^{})\propto
(k/k_{0})^{2}(1+z_{k}^{})^{-2}$.  For modes that re-enter the horizon
during radiation domination, when the $w_{r}^{}=1/3$ term dominates
the sum (\ref{def_F}), we solve (\ref{zk_Friedmann}) to find
$(1+z_{k}^{})\propto(k/k_{0})$, which implies the approximate scalings
$h_{c}^{}(k,\tau_{0})\propto (k/k_{0})^{-1}$ and
$\Omega_{gw}(k,\tau_{0})\propto(k/k_{0})^{0}$.  For modes that
re-enter during matter domination, when the $w_{m}^{}=0$ term
dominates the sum (\ref{def_F}), we find $(1+z_{k})\propto
(k/k_{0})^{2}$, which implies that $h_{c}^{}(k,\tau_{0})$ and
$\Omega_{gw}^{}(k,\tau_{0})$ are both $\propto (k/k_{0}^{})^{-2}$ in
this regime.

For a more proper analysis, consider a universe with 4 components:
matter ($w_{m}^{}=0$), curvature ($w_{K}^{}=-1/3$), dark energy
($w_{de}^{}(z)$), and radiation ($w_{r}^{}(z)=1/3+\delta
w_{r}^{}(z)$).  Note that, although one often assumes $w_{r}^{}=1/3$
during radiation domination, we have allowed for corrections $\delta
w_{r}^{}(z)$ due to early-universe effects discussed in subsection
\ref{EOS_corrections}.  Then we can write
\begin{equation}
  \label{F=Fhat+deltaF}
  F(z)=\hat{F}(z)+\delta F(z),
\end{equation}
where
\begin{equation}
  \label{def_Fhat}
  \hat{F}(z)=\Omega_{r}^{(0)}(1+z)^{2}+\Omega_{m}^{(0)}(1+z)
  +\Omega_{K}^{(0)},
\end{equation}
and
\begin{eqnarray}
  \delta F(z)\!&\!=\!&\!\Omega_{de}^{(0)}(1\!+\!z){\rm exp}\Big[
  3\!\!\int_{0}^{z}\!\!\!d\tilde{z}\frac{w_{de}^{}(\tilde{z})}
  {1\!+\!\tilde{z}}\Big] \nonumber\\
  &\!+\!&\!\Omega_{r}^{(0)}(1\!+\!z)^{2}\Big\{
  {\rm exp}\Big[3\!\!\int_{0}^{z}\!\!\!d\tilde{z}
  \frac{\delta w_{r}^{}(\tilde{z})}{1\!+\!\tilde{z}}\Big]\!-\!1\Big\}.
\end{eqnarray}
Here $\hat{F}$ represents a universe with spatial curvature, matter,
and ``standard'' ($w_{r}^{}=1/3$) radiation; and $\delta F$ contains
the modifications due to dark energy ($w_{de}^{}$) and
equation-of-state corrections ($\delta w_{r}^{}$).
    
If we neglect these modifications [by setting $\Omega_{de}^{(0)}=0
=\delta w_{r}^{}(z)$ so that $\delta F=0$], Eq.\ (\ref{zk_Friedmann})
has the exact solution
\begin{equation}
  \label{zhatk_soln}
  1+\hat{z}_{k}^{}\equiv\frac{1+z_{eq}^{}}{2}\left[-1+\sqrt{1+
      \frac{4[(k/k_{0})^{2}-\Omega_{K}^{(0)}]}{(1+z_{eq}^{})
        \Omega_{m}^{(0)}}}\;\right],
\end{equation}
where $1+z_{eq}^{}\equiv\Omega_{m}^{(0)}/\Omega_{r}^{(0)}$ is the
redshift of matter-radiation equality.  Then, including both
modifications, the solution becomes
\begin{equation}
  (1+z_{k}^{})=(1+\hat{z}_{k}^{})+\delta z_{k}^{}
\end{equation}
where $\hat{z}_{k}^{}$ is defined by Eq.\ (\ref{zhatk_soln}) and
\begin{equation}
  \label{delta_zk_2nd_order}
  \delta z_{k}^{}=\frac{F'(\hat{z}_{k}^{})}{F''(\hat{z}_{k}^{})}
  \left[-1+\sqrt{1-2\frac{F''(\hat{z}_{k}^{})\delta 
        F(\hat{z}_{k}^{})}{[F'(\hat{z}_{k}^{})]^{2}}}\;\right],
\end{equation}
This solution is obtained by Taylor expanding $F(z_{k}^{})$ around
$\hat{z}_{k}^{}$ (to 2nd order in $\delta z_{k}^{}=z_{k}^{}
-\hat{z}_{k}^{}$), and then solving the equation $F(z_{k}^{})=\hat{F}
(\hat{z}_{k}^{})$ for $\delta z_{k}^{}$.  It is extremely accurate for
a wide range of $\Omega_{de}^{(0)}$, $w_{de}^{}(z)$, and $\delta
w_{r}^{}(z)$.  Indeed, the simpler 1st-order expression
\begin{equation}
  \label{delta_zk_1st_order}
  \delta z_{k}^{}=-\frac{\delta F(\hat{z}_{k}^{})}{F'(\hat{z}_{k}^{})}
\end{equation}
is often sufficiently accurate.

\subsection{The horizon-crossing factor, $C_{2}$}
\label{horizon_thickness}

In the previous subsection, we treated horizon re-entry as a sudden
event that occurs when $k=aH$.  In reality, the ``outside'' behavior
($h_{k}={\rm constant}$) only holds when $k\ll aH$, and the ``inside''
behavior ($h_{k}\propto a^{-1}{\rm cos}[k\tau +{\rm phase}]$) only
holds when $k\gg aH$.  In between, when $k\sim aH$, neither behavior
holds --- {\it i.e.}, the horizon has a non-zero ``thickness.''

The behavior of the background equation-of-state
$w(\tau)=p(\tau)/\rho(\tau)$ during the period of horizon re-entry is
imprinted in the factor $C_{2}$.  For example, let us compute $C_{2}$
for a mode $k$ that re-enters the Hubble horizon when $w(\tau)$ is
varying slowly relative to the instantaneous Hubble rate.  Then we can
write
\begin{equation}
  a=a_{0}^{}(\tau/\tau_{0}^{})^{\alpha}\quad{\rm with}\quad 
  \alpha=\frac{2}{1+3w},
\end{equation}
so that the equation of motion for $\widetilde{h}_{k}$
\begin{equation}
  \widetilde{h}_{k}''+2(a'/a)\widetilde{h}_{k}'
  +k^{2}\widetilde{h}_{k}^{}=0
\end{equation}
has the solution
\begin{equation}
  \widetilde{h}_{k}(\tau)=h_{k}(\tau_{i}^{})\Gamma(\alpha+1/2)
  [k\tau/2]_{}^{1/2-\alpha}J_{\alpha-1/2}^{}(k\tau),
\end{equation}
where we have used $k\tau_{i}^{}\!\ll\!1$ and
$\widetilde{h}_{k}'(\tau_{i}^{}) \!=\!0$.  (Early in the radiation
era, the relevant modes were far outside the horizon, and the
corresponding mode functions were $\tau$-independent.)  We have
neglected the spatial curvature, $K$, because the two conditions $K\ll
a_{0}^{2}H_{0}^{2}$ (current observations indicate that the spatial
curvature is small) and $k>a_{0}H_{0}$ (we are only interested in
modes that are already inside the horizon) imply that $K$ produces a
negligible correction to the equation of motion for $h_{k}$.

Once the modes are well inside the horizon ($k\tau\gg 1$), we can use
the asymptotic Bessel formula
\begin{equation}
  J_{\alpha-1/2}(k\tau)\to\sqrt{\frac{2}{\pi k\tau}}\;
  {\rm cos}(k\tau-\alpha\pi/2)
\end{equation}
to find
\begin{equation}
  \label{Th_exact_alpha}
  \frac{\widetilde{h}_{k}^{2}(\tau)}{h_{k}^{2}(\tau_{i}^{})}=
  \frac{\Gamma^{2}(\alpha+1/2)}{\pi}[k\tau/2]^{-2\alpha}{\rm cos}^{2}
  (k\tau-\alpha\pi/2).
\end{equation}
On the other hand, since a mode $k$ re-enters the horizon
($k=aH=a'/a$) at time $\tau_{k}=\alpha/k$, we can rewrite Eq.\ 
(\ref{C1}) for $C_{1}$ as
\begin{subequations}
  \begin{eqnarray}
    C_{1}\!&\!=\!&\![\tau/\tau_{k}]_{}^{2\alpha}
    {\rm cos}^{2}[k(\tau-\tau_{k})+\phi_{k}^{}],\quad \\
    \label{Th_thin_alpha}
    \!&\!=\!&\![k\tau/\alpha]^{-2\alpha}{\rm cos}^{2}
    (k\tau-\alpha+\phi_{k}^{}).\quad
  \end{eqnarray}
\end{subequations}
Comparing Eqs.\ (\ref{def_C1}), (\ref{def_C2}), (\ref{Th_exact_alpha})
and (\ref{Th_thin_alpha}), we see that the phase shift $\phi_{k}^{}$
in Eq.\ (\ref{C1}) is given by
\begin{equation}
  \label{phi_k}
  \phi_{k}^{}=[1-\pi/2]\alpha,
\end{equation}
and that $C_{2}$ is given by
\begin{equation}
  \label{C2}
  C_{2}(k)=\frac{\Gamma^{2}(\alpha+1/2)}{\pi}[2/\alpha]_{}^{2\alpha},
\end{equation}
where $\alpha$ should be evaluated at horizon re-entry ($k=aH$).  In
particular, note that
\begin{equation}
  \begin{array}{lll}
    w=\;\!0\;\;\;\Rightarrow \!\!&\!\!C_{2}(k)=\frac{9}{16}
    \!&\! \;\;{\rm and}\quad\phi_{k}^{}=2-\pi, \\
    w=\frac{1}{3}\;\;\;\Rightarrow\quad \!\!&\!\!C_{2}(k)=\;1
    \!&\! \;\;{\rm and}\quad\phi_{k}^{}=1-\pi/2.
  \end{array}
\end{equation}

\subsection{The anisotropic-stress damping factor, $C_{3}$}
\label{free_streaming}

In this subsection, we will include the effects of the anisotropic
stress term $\Pi_{k}$ on the right-hand side of the tensor mode
equation (\ref{h_eq_ft}).  A non-negligible tensor anisotropic stress
$\Pi_{k}$ is most naturally generated by relativistic particles
free-streaming along geodesics that are perturbed by the presence of
tensor metric perturbations $h_{k}^{}$.  In this situation, Weinberg
\cite{Weinberg} has recently shown that the tensor mode equation
(\ref{h_eq_ft}) may be rewritten as a fairly simple
integro-differential equation for $h_{k}^{}$ --- see Eq.\ (18) in
\cite{Weinberg}.

Let us focus on a particularly interesting case: a radiation-dominated
universe in which the free-streaming particles constitute a
nearly-constant fraction $f$ of the background (critical) energy
density.  (Physically, if the free-streaming particles are stable, or
long-lived relative to the instantaneous Hubble time at re-entry, then
$f$ will indeed be nearly-constant, as required.)  In this case,
following an approach that is essentially identical to the one
outlined in \cite{DicusRepko}, we write the solution in the form
\begin{equation}
  h_{k}(\tau)=h_{k}(\tau_{i}^{})\sum_{n=0}^{\infty}a_{n}^{}j_{n}^{}(k\tau),
\end{equation}
where $j_{n}^{}(k\tau)$ are spherical Bessel functions, and find the
first five non-vanishing coefficients to be given by
\begin{subequations}
  \begin{eqnarray}
    a_{0}^{}\!&\!\!=\!\!&\!1, \\
    a_{2}^{}\!&\!\!=\!\!&\!\frac{10f}{(15+4f)}, \\
    a_{4}^{}\!&\!\!=\!\!&\!\frac{18f(3f+5)}{(15+4f)(50+4f)}, \\
    a_{6}^{}\!&\!\!=\!\!&\!\frac{\frac{130}{7}f(14f^{2}+92f+35)}
    {(15+4f)(50+4f)(105+4f)}, \\
    a_{8}^{}\!&\!\!=\!\!&\!\frac{\frac{85}{343}f(4802f^{3}\!+\!78266f^{2}
      \!+\!161525f\!-\!29400)}{(15+4f)(50+4f)(105+4f)(180+4f)\;\;\;}.\qquad
  \end{eqnarray}
\end{subequations}
The odd coefficients all vanish: $a_{2n+1}=0$.  Keeping these first
five non-vanishing terms yields a solution for $h_{k}^{}(\tau)$ that
is accurate to within $0.1\%$ for all values $0\!<\!f\!<\!1$.  Next,
as observed in \cite{DicusRepko}, we can use the asymptotic expression
\begin{equation}
  j_{2n}(k\tau)\to (-1)^{n}\frac{{\rm sin}(k\tau)}{k\tau}\quad{\rm as}
  \quad k\tau\to\infty,
\end{equation}
along with the $f=0$ solution $\widetilde{h}_{k}(\tau)=h_{k}
(\tau_{i}^{})j_{0} (k\tau)$ to infer that the tensor anisotropic
stress $\Pi_{k}$ induces no additional phase shift in $h_{k}$, so that
our earlier expression (\ref{phi_k}) for $\phi_{k}^{}$ is unchanged.
(See \cite{Bashinsky} for a complementary explanation of this null
result, based on causality.)  In this way, one also sees that
$\Pi_{k}$ damps the tensor power spectrum by the asymptotic factor
\begin{equation}
  C_{3}=|A|^{2},
\end{equation}
where
\begin{equation}
  A=\sum_{n=0}^{\infty}(-1)^{n}a_{2n}^{}.
\end{equation}
For example, keeping the first 4 terms in this sum, we find an
approximate expression for $A$:
\begin{equation}
  \frac{-\frac{10}{7}(98f^{3}-589f^{2}+9380f-55125)}
  {(15+4f)(50+4f)(105+4f)},
\end{equation}
which is accurate to within $1\%$ for all values $0\!<\!f\!<\!1$.  If
we keep the first 5 terms in the sum, we find an even better
approximation for $A$:
\begin{equation}
  \frac{15(14406f_{}^{4}\!-\!55770f_{}^{3}\!+\!3152975f_{}^{2}\!-\!
    48118000f\!+\!324135000)}{343(15+4f)(50+4f)(105+4f)(180+4f)}
\end{equation}
which is accurate to within $0.1\%$ for all values $0\!<\!f\!<\!1$.
These calculations improve on the accuracy of previous calculations
\cite{Bashinsky,PritchKam}.

\begin{figure}
  \begin{center}
    \includegraphics[width=3.1in]{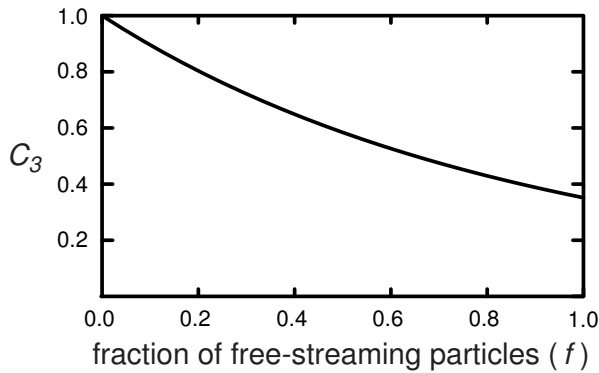}
  \end{center}
  \caption{$C_{3}$ is the transfer function factor that accounts for 
    the damping of the tensor power spectrum due to tensor anisotropic
    stress.  The factor depends on the fraction $f$ of the background
    (critical) energy density contained in free-streaming relativistic
    particles.  The figure plots this dependence for $0<f<1$.}
  \label{fsfig}
\end{figure}

The exact dependence of $C_{3}^{}$ on $f$ is shown in Fig.\ 
\ref{fsfig}.  Note that, as $f$ varies between $0$ and $1$, the
damping factor $C_{3}$ varies between $1.0$ and $0.35$.  In
particular, if we substitute $f=0.4052$, corresponding to 3 standard
neutrino species,the damping factor $0.80313$ agrees with the results
of numerical integrations \cite{Weinberg,PritchKam}.  When the modes
probed by the CMB re-enter the horizon, the temperature is relatively
low (corresponding to atomic-physics energies), so we are fairly
confident that neutrinos are the only free-streaming relativistic
particles.  But when the modes probed by laser interferometers
re-enter the horizon, the temperature is much higher (above the
electroweak phase transition, $T\sim10^{7}$~GeV), so that the physics
(and, in particular, the instantaneous free-streaming fraction $f$) is
much more uncertain.  Thus, laser interferometers offer the
possibility of learning about the free-streaming fraction $f$ in the
very early universe at temperatures between the inflationary and
electroweak symmetry breaking scales.

Finally, although Weinberg and subsequent authors have concentrated on
the tensor anisotropic stress due to a single fermionic species (the
neutrino), it is straightforward to generalize the analysis to include
a combination of species which ({\it i}) may each decouple at a
different time and temperature, and ({\it ii}) may be an arbitrary
mixture of bosons and fermions.  We find that, as long as all of these
free-streaming species decouple well before the modes of interest
re-enter the horizon, then all of the results presented in this
section are completely unchanged.  In other words, in order to
determine the behavior of the tensor mode function, one only needs to
know one number --- the total fraction $f$ of the critical density
contained in free-streaming particles --- even if the particles are a
mixture of fermionic and bosonic species with different temperatures
and decoupling times.

\subsection{Equation-of-state corrections, $\delta w_{r}^{}$}
\label{EOS_corrections}

In this section, we consider various physical effects that cause the
equation of state $w_{r}(z)$ to deviate from $1/3$ during the
radiation-dominated epoch, and the corresponding modifications that
these effects induce in the GWB transfer function.  Some of these
effects have been discussed previously by Seto and Yokoyama
\cite{SetoYokoyama}.  The deviations
\begin{equation}
  \delta w_{r}^{}(z)=w_{r}^{}(z)-1/3
\end{equation}
primarily modify the transfer function through the redshift factor
$(1+z_{k}^{})$ that appears in $C_{1}$ [see Eqs.\ 
(\ref{delta_zk_2nd_order}) and (\ref{delta_zk_1st_order})]; through
the horizon-crossing factor $C_{2}(k)$ [see Eq.\ (\ref{C2})]; and
through the phase shift $\phi_{k}^{}$ [see Eq.\ (\ref{phi_k})].  We
consider here six physical effects which can produce these kinds of
modifications of the transfer function.

First, deviations can be caused by mass thresholds in the early
universe.  Suppose that all particle species are described by
equilibrium distribution functions.  Then we can write $\rho$ and $p$
as
\begin{subequations}
  \begin{eqnarray}
    \rho&=&\frac{1}{2\pi^{2}}\sum_{i}g_{i}^{}T_{i}^{4}
    \int_{x_{i}^{}}^{\infty}\frac{(u^{2}-x_{i}^{2})^{1/2}}
    {{\rm exp}[u-y_{i}^{}]\pm1}u^{2}du, \\
    p&=&\frac{1}{6\pi^{2}}\sum_{i}g_{i}^{}T_{i}^{4}
    \int_{x_{i}^{}}^{\infty}\frac{(u^{2}-x_{i}^{2})^{3/2}}
    {{\rm exp}[u-y_{i}^{}]\pm1}du,
  \end{eqnarray}
\end{subequations}
where the $i$th species (with mass $m_{i}^{}$, and $g_{i}^{}$ internal
degrees of freedom) is described by temperature $T_{i}^{}$ and
chemical potential $\mu_{i}$, and we have defined the dimensionless
quantities $x_{i}^{}\equiv m_{i}^{}/T_{i}^{}$ and $y_{i}^{}\equiv
\mu_{i}^{}/T_{i}^{}$ \cite{KolbTurner}.  In the denominator, the $+$
and $-$ signs are for fermions and bosons, respectively.  Then the
deviation $\delta w_{r}^{}$ is given by the exact expression
\begin{equation}
  \delta w_{r}^{}=\sum_{i}\delta w_{r}^{(i)}
\end{equation}
where
\begin{equation}
  \delta w_{r}^{(i)}=-\frac{5}{\pi^{4}}\frac{g_{i}^{}}
  {g_{\ast\rho}^{}}\frac{T_{i}^{4}}{T_{}^{4}}f(x_{i}^{},y_{i}^{})
\end{equation}
represents the contribution from the $i$th species,
\begin{equation}
  g_{\ast\rho}^{}\equiv\sum_{i}g_{i}^{}\frac{T_{i}^{4}}{T_{}^{4}}
  \frac{15}{\pi^{4}}\int_{x_{i}^{}}^{\infty}
  \frac{(u^{2}-x_{i}^{2})^{1/2}u^{2}}{{\rm exp}[u-y_{i}]\pm1}du
\end{equation}
represents the effective number of relativistic degrees of freedom,
$T$ is conventionally chosen to be the photon temperature, and we have
defined the function
\begin{equation}
  f(x,y)\equiv x_{}^{2}\int_{x}^{\infty}\frac{(u^{2}
    -x_{}^{2})^{1/2}}{{\rm exp}[u-y]\pm1}du.
\end{equation}
For fixed $y_{i}^{}$, note that $f(x_{i}^{},y_{i}^{})$ vanishes as
$x_{i}^{}$ goes to $0$ or $\infty$; and in between it has a fairly
broad peak, with a maximum located at $x_{i}^{peak}$, and a peak value
$f_{i}^{peak}=f(x_{i}^{peak},y_{i})$.  In particular, when $y_{i}=0$,
then the ordered pair $(x_{i}^{peak}, f_{i}^{peak})$ is $(2.303,
1.196)$ for bosons and $(2.454, 1.125)$ for fermions.  This makes
sense: we expect $\delta w_{r}^{(i)}$ to vanish when $x_{i}^{}\ll 1$
(since the species is relativistic) and when $x_{i}^{}\gg 1$ (since
the species is non-relativistic, and makes a negligible contribution
to the energy density).  In between, when $x_{i}^{}\sim
x_{i,peak}^{}$, the $i$th species is cold enough to exhibit
non-relativistic behavior, yet hot enough to contribute non-negligibly
to the energy density.

Using the above equations, we can compute $\delta w_{r}^{}(z)$ once we
know $T_{i}^{}(z)$ and $\mu_{i}^{}(z)$.  But let us estimate the size
of the effect.  As a species becomes non-relativistic, it produces a
maximum equation-of-state deviation
\begin{equation}
  \delta w_{r}^{(i)}=-\frac{5f_{i}^{peak}}{\pi^{4}}
  \frac{g_{i}^{}}{g_{\ast\rho}^{}}\frac{T_{i}^{4}}{T{}^{4}}
\end{equation}
in the background equation of state.  Furthermore, if $N_{s}$
different species (with the same temperature and similar masses)
become non-relativistic at the same time, then (roughly speaking) the
effect is multiplied by $N_{s}$ (since their $\delta w_{r}^{(i)}$'s
add).  Ultimately, the fractional correction $\delta w_{r}^{}/
w_{r}^{}$ is model-dependent, but it can conceivably be as large as a
few percent.

Second, deviations can be produced by a trace anomaly in the early
universe.  During the radiation-dominated epoch, the universe is
dominated by highly relativistic particles whose masses may be
neglected.  Thus, each species is governed by a classical action that
is conformally invariant at the classical level, leading to the usual
conclusion that the stress-energy tensor is traceless and $w_{r}=1/3$.
But conformal invariance is broken at the quantum level by
interactions among the particles, so that $T_{\mu}^{\mu}\neq0$.  For
example, for a quark-gluon plasma governed by SU($N_{c}^{}$) gauge
theory, with $N_{f}^{}$ flavors, and gauge-coupling $g$, the equation
of state correction is given (up to ${\cal O}(g^{5})$ corrections) by
\cite{SU(Nc),grav_baryogen}
\begin{equation}
  \label{trace_anom}
  \delta w_{r}^{}=\frac{5}{18\pi^{2}}\frac{g^{4}}{(4\pi)^{2}}
  \frac{(N_{c}^{}+\frac{5}{4}N_{f}^{})(\frac{2}{3}N_{f}^{}-\frac{11}
    {3}N_{c}^{})}{2+\frac{7}{2}[N_{c}^{}N_{f}^{}/(N_{c}^{2}-1)]}.
\end{equation}
Note that this effect can be non-negligible: for large gauge groups
({\it i.e.}\ large $N_{c}^{}$) in the early universe (prior to the
electroweak phase transition), the equation of state may $w_{r}^{}$
may easily be reduced from $1/3$ by several percent, or more.

Third, deviations can be produced if the early universe behaves like a
slightly imperfect fluid.  The stress-energy tensor for an imperfect
fluid contains (in addition to the usual perfect-fluid terms) three
extra terms whose coefficients ($\chi$, $\eta$, and $\zeta$) represent
heat conduction, shear viscosity, and bulk viscosity (see Weinberg
\cite{WeinbergGR}, Ch.\ 2.11).  Of these dissipative effects, only the
bulk viscosity term
\begin{equation}
  \Delta T^{\mu\nu}=-\zeta(g^{\mu\nu}+U^{\mu}U^{\nu})
  U^{\lambda}_{\;\;;\lambda}
\end{equation}
can contribute to the background evolution in an FRW universe (see
Weinberg \cite{WeinbergGR}, Chs.\ 15.10-15.11).  This term modifies
the continuity equation
\begin{equation}
  \dot{\rho}=-3H(\rho+p)+9\zeta H^{2}=-3H\rho\Big[1+w-\frac{8\pi G
    \zeta}{H}\Big]
\end{equation}
so that, as far as gravitational waves are concerned, the effective
equation is corrected by
\begin{equation}
  \delta w_{r}^{}=-\frac{8\pi G\zeta}{H}.
\end{equation}

Whereas the three effects discussed thus far produce small corrections
to the equation of state, it is worth mentioning three other effects
that can produce much larger deviations.  The first example is a
massive particle species that decouples from the thermal plasma before
its abundance becomes negligible.  Since its energy density falls as
$a^{-3}$ (more slowly than the radiation density, which falls as
$a^{-4}$), it can come to dominate the energy density of the universe
before it decays (if its lifetime $\tau_{decay}$ is sufficiently
long).  In this case, $w$ drops to zero when the particle becomes
dominant, and rises back to $w=1/3$ over a timescale given by the
decay lifetime $\tau_{decay}$.  Second, extra-dimensional physics
typically modifies the effective 4-dimensional Friedmann equation.
Such modifications which, from the standpoint of the GWB, can in some
cases look like corrections to the effective radiation equation of
state, are primarily constrained by the requirement that the Friedmann
equation becomes sufficiently similar to the ordinary 4-dimensional
Friedmann equation (with ordinary matter) by the time of BBN.  Third,
due to the weak coupling of the inflaton, the temperature at the start
of the radiation dominated epoch (the reheat temperature) can be much
lower than the energy scale at the end of inflation.  In this case,
laser interferometer scales might actually re-enter the horizon during
the reheating epoch (before the start of radiation domination), when
the equation of state was probably quite different from $w=1/3$.  The
actual equation-of-state depends on the details of the reheating
process, but a commonly-discussed value is $w=0$, or some value in the
range $0<w<1/3$ \cite{reheating}.  If $w=0$ during reheating, the
corresponding modification of the GWB might be similar to the
modification due to the long-lived massive relic discussed above.

Note that the first of these six processes can be expressed as a
modification of $g_{\ast}$, and the effects on the transfer function
can be computed using the methods discussed in
Ref.~\cite{SmithKamCoor}.  However, the other five cannot.

\section{Discussion}
\label{discussion}

\begin{figure}
  \begin{center}
    \includegraphics[width=3.1in]{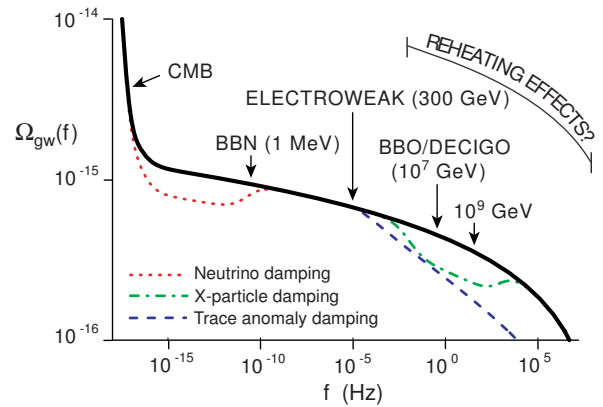}
  \end{center}
  \caption{The black solid curve represents the present-day 
    gravitational-wave energy spectrum, $\Omega_{gw}^{}(f,\tau_{0})$,
    for the inflationary model $V(\phi)=(1/2)m^{2}\phi^{2}$.  The red
    dotted curve shows the damping effect due to (three ordinary
    massless species of) free-streaming neutrinos.  The green
    dot-dashed curve shows the damping effect which arises if
    free-streaming particles make up fifty percent of the background
    energy density at the time $\tau_{BBO}^{}$ when the modes probed
    by BBO/DECIGO re-enter the horizon.  As shown in the figure, the
    particles begin free-streaming sometime before $\tau_{BBO}^{}$,
    and decay sometime after $\tau_{BBO}^{}$, but prior to electroweak
    symmetry breaking.  Finally, the blue dashed curve shows the
    effect of a conformal anomaly in the early universe that slightly
    reduces the equation of state from $w=0.33$ to $w=0.31$ above the
    electroweak phase transition.  The spectrum will also be modified
    on comoving scales that re-enter the horizon during the reheating
    epoch after inflation; but the range of scales affected by
    reheating is unknown.  Finally, note that the correlated BBO
    interferometer proposal claims a sensitivity that extends beyond
    the bottom of the figure (down to roughly
    $\Omega_{gw}^{}\sim10^{-17}$) in the frequency range from
    $10^{-1}$~Hz to $10^{0}$~Hz.}
  \label{energy_plot}
\end{figure}

Fig.~\ref{energy_plot} illustrates some of the effects discussed in
section \ref{transfer_fn}.  In this figure, the solid black curve
represents the present-day energy spectrum, $\Omega_{gw}(f,
\tau_{0}^{})$, generated by a particular inflationary model ---
namely, a quadratic potential $V(\phi)=(1/2)m^{2}\phi^{2}$.  The red
dotted curve illustrates the damping due to tensor anisotropic stress
from free-streaming neutrinos.  We have assumed that the
free-streaming fraction is $f=0.4052$, which is the $f$ value for
three standard neutrino species which decouple around the time of BBN.
The green dot-dashed curve represents the damping due to tensor
anisotropic stress from various particle species ($X$ particles) which
begin free-streaming before the scales detected by BBO/DECIGO re-enter
the horizon and then decay after the scales re-enter, but prior to the
electroweak phase transition.  As an example, we have assumed that the
free-streaming fraction is $f=0.5$.  Finally, the blue dashed curve
represents damping due to a trace anomaly that is present above the
electroweak scale.  For illustration, we have assumed that this
anomaly, through Eq.\ (\ref{trace_anom}), reduces the equation of
state from $w_{r}^{}=1/3$ by $\delta w_{r}^{}=-0.02$.  This reduction
may be achieved by various combinations of the number of colors
$N_{c}^{}$, the number of flavors $N_{f}^{}$, and the gauge coupling
$g$; but the point is that we have not chosen an unreasonable large
value for $\delta w_{r}^{}$, given the large gauge groups that are
often theorized to be present at high energies.

The key point conveyed by Fig.~\ref{energy_plot} is that there are a
variety of plausible post-inflationary effects that can produce rather
large modifications of the gravitational-wave spectrum on
laser-interferometer scales, without modifying the spectrum on CMB
scales.  This is tantalizing, since the modifications on
laser-interferometer scales reflect the primordial dark age between
the end of inflation and the electroweak phase transition, at energies
beyond the reach of terrestrial particle accelerators.

\begin{figure}
  \begin{center}
    \includegraphics[width=3.1in]{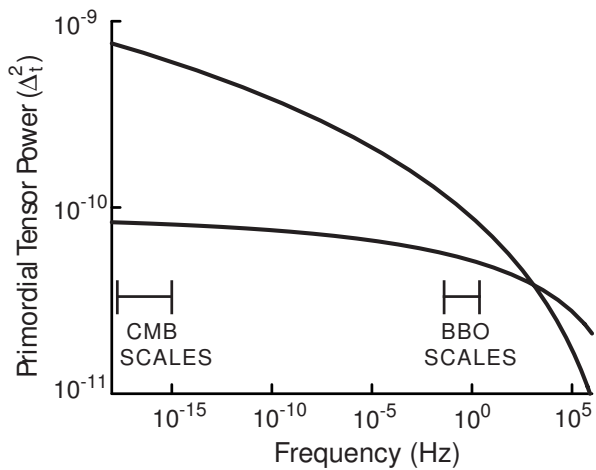}
  \end{center}
  \caption{The precise predictions of the primordial tensor power 
    spectrum for two inflationary models whose parameters have been
    chosen so that they make identical predictions for the scalar
    fluctuation amplitude and spectral index on CMB scales.  The two
    models are the quartic monomial potential
    $V(\phi)=\lambda\phi^{4}$ (top), and the axion (or ``natural
    inflation'') potential $V(\phi)=M_{I}^{4} [1+{\rm
      cos}(\phi/\mu)]$.}
  \label{primordial}
\end{figure}

Let us look ahead to future observations.  First, as measurements of
the scalar power spectrum improve, we will be restricted to
considering inflationary models that match precisely the measured
scalar perturbation spectrum on CMB scales.  This is not enough to
specify a unique model because there is a range of inflaton potentials
whose scalar perturbation spectra are indistinguishable on CMB scales.
For example, two such models are shown in Fig.~\ref{primordial}.
The two curves correspond to two different inflationary models whose
parameters have been chosen so that they produce the same scalar
amplitude and tilt on CMB scales, and match the observed scalar
amplitude.  From the figure, though, we see that their primordial
tensor power spectra are distinguishable.  The upper solid curve
corresponds to the simple quartic monomial potential
$V(\phi)=\lambda\phi^{4}$, and the lower solid curve corresponds to
the axion (or ``natural inflation'' \cite{NaturalInflation}) potential
$V(\phi)=M_{I}^{4}[1+{\rm cos}(\phi/\mu)]$.

Note that the two spectra ``curve'' significantly downward over the
broad range of wavenumbers shown in Fig.~\ref{primordial}, so that
they are not well described by a power-law (straight-line)
approximation over this range.  In particular, a power-law
extrapolation based on CMB measurements will give a fairly poor
estimate of the real spectrum on laser-interferometer scales.  In
fact, since the real spectra tend to redden increasingly on smaller
scales ({\it i.e.}\ as the end of inflation approaches), the CMB
power-law extrapolation will tend to overestimate the tensor spectrum
on laser-interferometer scales.

Second, note that the inflationary consistency relation
(\ref{consistency}) implies that a gravitational-wave spectrum with a
larger tensor/scalar ratio $r$ on CMB scales also tends to have a
``redder'' (more negative) tilt $n_{t}$.  For gravitational waves with
detectably-large amplitudes on CMB scales, this leads to the
``convergence effect'' in Fig.~\ref{primordial}: if two different
inflationary models produce scalar perturbations with the same
amplitude and tilt on large (CMB) scales, then their tensor spectra
tend to approach and even cross one another on small
(laser-interferometer) scales, as illustrated by the two solid black
curves in Fig.~\ref{primordial}.  As a result, once one has measured
the scalar and tensor perturbations on CMB scales, one expects the
tensor spectrum on laser-interferometer scales to lie within a
relatively narrow and predictable range.

This is both good news and bad news for laser-interferometer science.
The bad news is that the convergence effect tends to make laser
interferometers less effective at breaking any remaining degeneracies
between inflationary models because their predictions for the tensor
spectrum are so similar on small scales.  The degree to which laser
interferometers can break such degeneracies has been studied in
\cite{SmithKamCoor}.  The good news is that the measurements can
potentially provide a model-independent test of a key inflationary
prediction: that the tensor spectrum observed at long wavelengths
really does extend to small wavelengths, as expected from the
consistency relation (\ref{consistency}).  Furthermore, precise
laser-interferometer measurements can serve as sensitive probes of new
post-inflationary physics that modifies the transfer function on small
scales.

In this light, let us now imagine some time in the future in which we
have successfully measured the tensor amplitude on CMB scales, and
consider four possible outcomes of a laser-interferometer experiment
(like BBO).  First, suppose that BBO detects the gravitational-wave
amplitude at the expected value, as exemplified in
Fig.~\ref{primordial}.  This would be a quantitative confirmation of
inflation generally, and of the inflationary consistency condition
(\ref{consistency}) in particular, since the large difference in
wavenumber between CMB and laser-interferometer scales provides a
large lever-arm to measure the tensor tilt $n_{t}^{}$.  It would also
suggest an upper bound on the size of any exotic effects in the
transfer function on laser-interferometer scales.

Second, suppose that BBO detects the gravitational-wave amplitude
significantly {\it below} its expected value.  Then the interpretation
is less straighforward.  On the one hand, we could interpret the
suppression as being due to transfer-function effects, such as those
discussed above.  With this interpretation, the suppression would
imply a rare opportunity to measure physical properties of the early
universe, at temperatures above the electroweak scale, when the
relevant modes re-entered the horizon.  On the other hand, we could
interpret the suppression to mean that inflation is more complicated
than we expected: perhaps there is a feature in the inflaton potential
which suppresses the primordial tensor spectrum on small scales
relative to our expectations, or perhaps the consistency condition
itself is violated (as it is in some multi-field models of inflation).
Further measurements of the tensor spectrum at intermediate scales are
needed to resolve this issue.

Third, suppose that BBO detects the gravitational-wave amplitude
significantly {\it above} its expected value.  At this point it is
worth noting that virtually all of the transfer-function effects
mentioned in section \ref{transfer_fn} (including tensor anisotropic
stress, the late decay of a massive relic species, bulk viscosity, a
conformal anomaly, or standard reheating with $w\leq 1/3$) {\it
  suppress} the gravitational-wave spectrum on small scales, relative
to large scales.  In order to {\it enhance} the spectrum on
laser-interferometer scales, one must invoke an even more exotic
effect, such as a reheating epoch with a low reheat temperature and an
unusual equation of state ($w>1/3$); or the production of
gravitational waves {\it after} inflation due to the collision of
bubbles from a phase transition; or perhaps extra-dimensional physics.

Fourth, suppose that BBO detects nothing at all.  This would most
likely indicate a fundamental problem with inflation, since an period
of accelerating expansion produces a broad and nearly-flat spectrum of
primordial tensor perturbations quite generically (regardless of the
particular model that drives inflation).  One way out of this
conclusion would be to invoke an extreme suppression effect in the
transfer function on BBO scales --- {\it e.g.} due to a reheating
epoch with equation of state $w=0$ and a reheat temperature well below
$10^{7}$~GeV.  Determining which interpretation is correct would
require digging deeper, with more sensitive experiments on BBO scales.

\end{document}